\documentclass[12pt]{iopart}

\usepackage[dvipdf]{epsfig}

\usepackage{graphicx}

\begin{document}
\title[]
{Conformal coupling associated with the Noether symmetry and its connection with the $\Lambda$CDM dynamics}

\author{Rudinei C. de Souza\footnote{rudijantsch@gmail.com}\ and Gilberto M. Kremer\footnote{kremer@fisica.ufpr.br}}

\address{\ Departamento de F\'\i sica, Universidade Federal do Paran\'a,
 Curitiba, Brazil}

\begin{abstract}
The aim of the present work is to investigate  a non-minimally coupled scalar field model through the Noether symmetry approach, with the radiation, matter and cosmological constant eras being analyzed. The Noether symmetry condition allows a conformal coupling and by means of a change of coordinates in the configuration space the field equations can be reduced to a single equation, which is of the form of the Friedmann equation for the $\Lambda$CDM model. In this way, it is formally shown that the dynamical system can furnish solutions with the same form as those of the $\Lambda$CDM model, although the theory here considered is physically different from the former. The conserved quantity associated with the Noether symmetry can be related to the kinetic term of the scalar field and could constrain the possible deviations of the model from the $\Lambda$CDM picture. Observational constraints on the variation of the gravitational constant can be imposed on the model through the initial condition of the scalar field.
\end{abstract}

 \pacs{98.80.-k, 98.80.Cq, 95.35.+d }
 \maketitle

\section{Introduction}

The observational astronomy evidences that the Universe is acceleratedly expanding at the present \cite{1}, a fact that is not expected from the standard cosmology. The first attempt to explain this unexpected occurrence was to revive the cosmological constant model. Unfortunately, such a model soon presented inconsistence problems \cite{2.1} and  new models appeared in the literature, but still none could be considered as the definitive model \cite{2.a}. Among these attempts, the most popular is the scalar field model \cite{3}, which can present several acceptable forms for the potentials \cite{5}. Still in this context, there are the non-minimally coupled scalar field models \cite{7.1} (their origin is in the Brans-Dicke theory \cite{7.2}), which present a new physics behind them -- a variable gravitational "constant". The models with fermion fields \cite{7.3}, tachyon fields \cite{7.4}, Chaplygin gas \cite{7.6}, van der Waals gas \cite{7a} and the $f(R)$ theories \cite{7.7} -- which generalizes the Einstein-Hilbert action -- are alternative to the scalar field model. The $f(R)$ theories are mathematically equivalent to the non-minimally coupled scalar field models \cite{7.8}.

Although this variety of alternative theories, the observational data that confront the General Relativity plus cosmological constant with alternative theories appear to conspire in behalf of the $\Lambda$CDM model \cite{8} -- at least from the matter dominated era until the present. Without ignoring such an observational fact, this work intends to investigate a general non-minimally coupled scalar field theory through the Noether symmetry approach (see e.g. \cite{7.7,9.0,9}), comprehending the radiation, matter and cosmological constant eras. The work \cite{14} studies the Noether symmetry in general extended gravity theories applied to cosmology. Although the non-minimally coupled scalar field model is physically different from the cosmological constant model, by means of a change of coordinates in the configuration space generated by the Noether symmetry, it is possible to obtain from the original field equations a dynamics that is essentially the same of that from the $\Lambda$CDM model.

The paper \cite{9.1} analyzes a Universe composed of a scalar field non-minimally coupled in the context of the inflationary era (\emph{inflaton} field) using the Noether symmetry approach, and several forms for the potential and coupling are obtained. Here this approach is employed for analyzing a non-minimally coupled scalar field in the context of a dark energy field, with the matter and radiation fields being present. This task takes us to a stronger restriction for the forms of the potential and coupling in order to have a Noether Lagrangian. Under our considerations, a unique set of functions that represent the potential and coupling is obtained, which furnishes the above mentioned dynamics.

The work is organized as follows: In section two the action of the model is presented, a point-Like Lagrangian is derived from it and the field equations are written. In the third section the Noether potential and coupling are determined. In section four the dynamical system is written in a new coordinate system and the field equations are solved for the known eras of the Universe. The fifth section closes the work with the main conclusions.

In this work we use the signature $(+,-,-,-)$, the natural units $8\pi G=\hbar=c=1$, and the Ricci tensor defined as $R_{\mu\nu}=R^\lambda_{\mu\nu\lambda}$, such that the Ricci scalar for a flat Friedmann-Robertson-Walker (F-R-W) metric has the form
\begin{eqnarray}
R=g^{\mu\nu}R_{\mu\nu}=6\left(\frac{\ddot a}{a}+\frac{\dot a^2}{a^2}\right).
\end{eqnarray}

\section{Field equations}

Let us take the following general action for a Universe composed of  matter and radiation fields and a scalar field non-minimally coupled to the gravity
\begin{eqnarray}
S=\int d^4x\sqrt{-g}\ \bigg\{L(\phi)R+\frac{1}{2}g^{\mu\nu}\partial_\mu\phi\partial_\nu\phi-U(\phi)+\mathcal{L}_m+\mathcal{L}_r\bigg\}, \label{ga}
\end{eqnarray}
where $\phi$ describes the scalar field, and $\mathcal{L}_m$ and $\mathcal{L}_r$ are the Lagrangian densities of the matter and radiation fields, respectively. Here $R$ denotes the Ricci scalar,  $L(\phi)$ the coupling with the gravity and $U(\phi)$  the self-interaction potential of the scalar field.

By varying the action (\ref{ga}) with respect to the metric tensor $g_{\mu\nu}$ we arrive at the modified Einstein field equations
\begin{equation}
R_{\mu\nu}-\frac{1}{2}g_{\mu\nu}R=-\frac{T_{\mu\nu}}{2L}. \label{efe}
\end{equation}
Above, $T_{\mu\nu}=T_{\mu\nu}^m+T_{\mu\nu}^r+T_{\mu\nu}^\phi$, where the letters $m, r$ and $\phi$ label the energy-momentum tensors of the matter, radiation and scalar fields, respectively. They are defined by
\begin{eqnarray}
T_{\mu\nu}^m=\frac{2}{\sqrt{-g}}\frac{\delta(\sqrt{-g}\mathcal{L}_m)}{\delta
g^{\mu\nu}},\quad T_{\mu\nu}^r=\frac{2}{\sqrt{-g}}\frac{\delta(\sqrt{-g}\mathcal{L}_r)}{\delta
g^{\mu\nu}}, \label{emt1}
\\
T_{\mu\nu}^\phi=\partial_\mu\phi\partial_\nu\phi-\bigg(\frac{1}{2}\partial_\theta\phi\partial^\theta\phi-
U\bigg)g_{\mu\nu}+2(\nabla_\mu\nabla_\nu-g_{\mu\nu}\nabla_\theta\nabla^\theta)L.\qquad\qquad \label{emt2}
\end{eqnarray}
Observe that when $L\longrightarrow 1/2$ in equation (\ref{efe}) the Einstein field equations are recovered. Considering that the matter and radiation fields do not interact with the scalar field, their energy-momentum tensors must be conserved, i.e. $\nabla_\mu T^{\mu\nu}_m=\nabla_\mu T^{\mu\nu}_r=0$. Due to the non-minimal coupling, the energy momentum tensor of the scalar field (\ref{emt2}) is not conserved as usual, its covariant derivative given by
\begin{equation}
\nabla_\mu T^{\mu\nu}_{\phi}=\frac{\partial_\mu L}{L}\left(T^{\mu\nu}_{m}+T^{\mu\nu}_{r}+T^{\mu\nu}_{\phi}\right).
\end{equation}

Now we consider a flat F-R-W metric, with a spatially homogeneous field $\phi$, a pressureless matter field  and a radiation field with the usual  equations of state $p_m=0$ and $p_r=\rho_r/3$, respectively. The conservation condition of the energy-momentum tensor of the matter and radiation fields, $\nabla_\mu T^{\mu\nu}_m=\nabla_\mu T^{\mu\nu}_r=0$, determines the evolution equations for the energy densities
\begin{equation}\label{ener}
\dot\rho_m+3\frac{\dot a}{a}\rho_m=0, \qquad \dot\rho_r+3\frac{\dot a}{a}(\rho_r+p_r)=0,
\end{equation}
where the point represents time derivative and $a$ is the scale factor.
Equations (\ref{ener}) imply the well-known solutions $\rho_m=\rho_m^0/a^3$ and $\rho_r=\rho_r^0/a^4$, with $\rho_m^0$ and $\rho_r^0$ denoting the energy densities of the matter and radiation at an initial instant. Using these energy densities, from (\ref{emt1}) we determine the respective Lagrangians
\begin{equation}
\mathcal{L}_m=-\rho_m=-\frac{\rho_m^0}{a^3}, \qquad\mathcal{L}_r=-\rho_r=-\frac{\rho_r^0}{a^4}.
\end{equation}

One keeping the above considerations and results, we get from action (\ref{ga}) by application of the flat F-R-W metric and integration by parts, through the elimination of the frontier terms, the point-like Lagrangian
\begin{equation}
\mathcal{L}=6a\dot{a}^2L+6a^2\dot a\frac{dL}{d\phi}\dot\phi-a^3\bigg(\frac{\dot\phi^2}{2}-U\bigg)+\rho_m^0+\frac{\rho_r^0}{a}, \label{plL}
\end{equation}
which furnishes the same dynamical equations as those from the field equations (\ref{efe}) for a flat F-R-W metric.

The modified Friedmann equation is obtained by imposing that the "energy function" associated with the Lagrangian (\ref{plL}) vanishes, i.e.
\begin{equation}
E_\mathcal{L}=\frac{\partial \mathcal{L}}{\partial \dot{a}}
\dot{a}+\frac{\partial \mathcal{L}}{\partial \dot{\phi}}
\dot{\phi}-\mathcal{L}=0\;
\Longrightarrow\;\bigg(\frac{\dot a}{a}\bigg)^2=\frac{\rho_m+\rho_r+\rho_\phi}{6L},\label{EF}
\end{equation}
with $\rho_\phi$ denoting the energy density of the scalar field.

From the Euler-Lagrange equation for $a$ and $\phi$, applied to (\ref{plL}), we obtain the acceleration and modified Klein-Gordon equations
\begin{eqnarray}
\frac{\ddot{a}}{a}=-\frac{[\rho_m+\rho_r+\rho_\phi+3(p_r+p_\phi)]}{12L}, \label{acce}\\
\ddot\phi+3\bigg(\frac{\dot a}{a}\bigg)\dot\phi-6\bigg[\frac{\ddot a}{a}+\bigg(\frac{\dot a}{a}\bigg)^2\bigg]\frac{dL}{d\phi}+\frac{dU}{d\phi}=0,
\end{eqnarray}
respectively, where $p_\phi$ is the pressure of the scalar field. In the derivation of (\ref{acce}), we substituted (\ref{EF}) in the term $\dot a^2/a^2$, such that the energy densities pass to appear in (\ref{acce}).

In the above equations, the energy density and the pressure of the scalar field are defined by
\begin{eqnarray}
\rho_\phi=\frac{1}{2}\dot\phi^2+U-6\bigg(\frac{\dot a}{a}\bigg)\frac{dL}{d\phi}\dot\phi, \label{ed}\\ p_\phi=\frac{1}{2}\dot\phi^2-U+2\Bigg[\frac{dL}{d\phi}\ddot\phi+2\bigg(\frac{\dot a}{a}\bigg)\frac{dL}{d\phi}\dot\phi
+\frac{d^2L}{d\phi^2}\dot\phi^2\Bigg], \label{pre}
\end{eqnarray}
in accordance with the energy-momentum tensor (\ref{emt2}).

\section{Noether functions}

The Noether symmetry approach can be used as a formal suggestion for the undefined potentials and
couplings of a Lagrangian. Once the undefined forms are determined from a Noether symmetry
condition, the dynamical system  presents additionally a conserved quantity and a cyclic variable is found through a change of coordinates in the configuration space, which can be useful for the integration of the system.

A Noether symmetry exists for a given Lagrangian of the form $\mathcal{L}=\mathcal{L}(q_k, \dot q_k)$ if the condition $L_\textbf{x}\mathcal{L}=0$ is satisfied. The $L_\textbf{x}$ denotes the Lie derivative with respect to the infinitesimal generator of symmetry $\textbf{X}$, defined by
\begin{equation}
\textbf{X}=\sum_k\Bigg(\alpha_k\frac{\partial}{\partial q_k}+
\frac{d\alpha_k}{dt}\frac{\partial}{\partial\dot q_k}\Bigg), \label{NS1}
\end{equation}
where the $\alpha_k$'s are functions of the generalized coordinates
$q_k$.

The conserved quantity (or constant of motion) associated with the Noether symmetry established by the above condition is given by
\begin{equation}
M_0=\sum_k\alpha_k\frac{\partial\mathcal{L}}{\partial \dot q_k}. \label{NS2}
\end{equation}

The new set of variables $\{Q_k(q_l)\}$ for the configuration space, such that one of the variables is cyclic, obeys the following system of differential equations
\begin{eqnarray}
\sum_l\Bigg(\alpha_l\frac{\partial u_{k'}}{\partial q_l}\Bigg)=0,\qquad
\sum_l\Bigg(\alpha_l\frac{\partial z}{\partial q_l}\Bigg)=1, \label{NS3}
\end{eqnarray}
where $k'=1, 2, ... \ k-1$ and the $u_{k-1}$ and $z$ comprehend the new coordinates $Q_k(q_l)$ of the configuration space, with $z$ being the cyclic variable.

Applying the Noether condition $L_\textbf{x}\mathcal{L}=0$ to (\ref{plL}), with $\textbf{X}$  defined for our problem as
\begin{equation}
\textbf{X}=\alpha\frac{\partial}{\partial a}+\beta\frac{\partial}{\partial \phi}+
\frac{d\alpha}{dt}\frac{\partial}{\partial\dot a}+\frac{d\beta}{dt}\frac{\partial}{\partial\dot \phi},
\end{equation}
where $\alpha$ and $\beta$ are functions of $a$ and $\phi$, we arrive at the following system of differential equations

\begin{eqnarray}
\alpha+2a\frac{\partial\alpha}{\partial a}+\frac{a}{L}\frac{dL}{d\phi}\bigg(\beta+a\frac{\partial\beta}{\partial a}\bigg)=0,\label{sys1}\\
3\alpha-12\frac{dL}{d\phi}\frac{\partial\alpha}{\partial\phi}+2a\frac{\partial\beta}{\partial \phi}=0,\label{sys2}
\\
a\beta\frac{d^2L}{d\phi^2}+\bigg(2\alpha+a\frac{\partial\alpha}{\partial a}+a\frac{\partial\beta}{\partial\phi}\bigg)\frac{dL}{d\phi}+2L\frac{\partial\alpha}{\partial \phi}-\frac{a^2}{6}\frac{\partial\beta}{\partial a}=0,\label{sys3}\\
\alpha\bigg(3U-\frac{\rho^0_r}{a^4}\bigg )+a\beta\frac{dU}{d\phi}=0. \label{sys4}
\end{eqnarray}

We are interested in solutions of the above system when $dL/d\phi\neq0$. Then, searching for $\alpha$ and $\beta$ that are given by separable functions of $a$ and $\phi$, i.e. $\alpha=\alpha_1(a)\alpha_2(\phi)$ and $\beta=\beta_1(a)\beta_2(\phi)$,
the system (\ref{sys1})-(\ref{sys4}) furnishes the solution
\begin{equation}
\alpha=0,\qquad \beta=\frac{\beta_0}{a}, \qquad U=\Lambda , \qquad L=A+B\phi-\frac{\phi^2}{12}, \label{symm_2}
\end{equation}
where $\beta_0$, $\Lambda$, $A$ and $B$ are constants. The same solution can also be obtained from equations (21)-(24) of \cite{9.0} by choosing $\alpha=0$.

Hence, with $L$ and $U$ given by (\ref{symm_2}), the system presents a Noether symmetry with the constant of motion given by
\begin{equation}
M_0=\alpha\frac{\partial \mathcal{L}}{\partial \dot a}+\beta\frac{\partial \mathcal{L}}
{\partial \dot\phi}=\beta_0a^2\bigg[6\bigg(\frac{\dot a}{a}\bigg)\bigg(B-\frac{\phi}{6}\bigg)-\dot\phi\bigg], \label{cq_2}
\end{equation}
which was determined from (\ref{NS2}).

\section{Analysis of the system}

From the found Noether potential $U=\Lambda$ and coupling  $L=A+B\phi-\frac{\phi^2}{12}$, we have the following dynamical system to solve
\begin{eqnarray}
6\bigg(A+B\phi-\frac{\phi^2}{12}\bigg)\frac{\dot a^2}{a^2}=\frac{\rho_m^0}{a^3}+\frac{\rho_r^0}{a^4}+\frac{\dot\phi^2}{2}+\Lambda-6\frac{\dot a}{a}\bigg(B-\frac{\phi}{6}\bigg)\dot\phi, \label{sysa}\\
\ddot\phi+3\bigg(\frac{\dot a}{a}\bigg)\dot\phi-6\bigg[\frac{\ddot a}{a}+\bigg(\frac{\dot a}{a}\bigg)^2\bigg]\bigg(B-\frac{\phi}{6}\bigg)=0,\label{sysb}\\
6\bigg(\frac{\dot a}{a}\bigg)\bigg(B-\frac{\phi}{6}\bigg)-\dot\phi=\frac{M_0}{\beta_0a^2}, \label{sysc}
\end{eqnarray}
including the additional equation from the constant of motion (\ref{cq_2}).

If we perform the transformation $\phi\longrightarrow\phi\ +$ constant and take the constant equal to $6B$, the action (\ref{ga}) with the above Noether forms (valid for the flat F-R-W metric)
\begin{eqnarray}
&S=\int d^4x\sqrt{-g}\ \left\{\left(A+B\phi-\frac{\phi^2}{12}\right)R+\frac{1}{2}g^{\mu\nu}\partial_\mu\phi\partial_\nu\phi-\Lambda+\mathcal{L}_{m}+\mathcal{L}_{r}\right\},\nonumber\\
\end{eqnarray}
transforms into
\begin{eqnarray}
S=\int d^4x\sqrt{-g}\ \left\{\left(A-3B^2-\frac{\phi^2}{12}\right)R+\frac{1}{2}g^{\mu\nu}\partial_\mu\phi\partial_\nu\phi-\Lambda+\mathcal{L}_{m}+\mathcal{L}_{r}\right\}.\nonumber\\ \label{ta}
\end{eqnarray}
We can set $A-3B^2=1/2$ in the transformed action and obtain the conformal coupling $L=\frac{1}{2}\Big(1-\frac{\phi^2}{6}\Big)$. This is equivalent to choose $A=1/2$ and $B=0$ in $L$ for the original variable without loss of generality. Having in mind the result from the transformation (\ref{ta}), one can return to the original problem with the function $L=\frac{1}{2}\Big(1-\frac{\phi^2}{6}\Big)$ as being the most general one, which connects the conformal coupling with a symmetry of the model. In so doing, our starting Lagrangian is expressed in the form
\begin{equation}
\mathcal{L}=3a\dot{a}^2-\frac{1}{2}a\dot{a}^2\phi^2-a^2\dot a\phi\dot\phi-a^3\bigg(\frac{\dot\phi^2}{2}-\Lambda \bigg)+\rho_m^0+\frac{\rho_r^0}{a}. \label{plL1}
\end{equation}

Since we have a Noether symmetry for the Lagrangian (\ref{plL1}), according to (\ref{NS3}) there is a transformation of variables that obeys the system
\begin{eqnarray}\label{t1}
\alpha\frac{\partial u}{\partial a}+\beta\frac{\partial u}{\partial \phi}=0,\qquad
\alpha\frac{\partial z}{\partial a}+\beta\frac{\partial z}{\partial \phi}=1,
\end{eqnarray}
where $z$ is the cyclic variable associated with the conserved quantity $M_0$. For the found set $(\alpha=0, \beta=\beta_0/a)$, the system (\ref{t1}) has the solution
\begin{equation}
u=f(a), \qquad z=\frac{a\phi}{\beta_0}+g(a). \label{s1,2}
\end{equation}
We choose from the general solution (\ref{s1,2}) the particular forms
$u=a$ and  $z=a\phi/\beta_0,$
which transform the Lagrangian (\ref{plL1}) into
\begin{equation}
\mathcal{L}=3a\dot{a}^2-\frac{\beta_0^2}{2}a\dot z^2+a^3\Lambda +\rho_m^0+\frac{\rho_r^0}{a}. \label{plL2}
\end{equation}

The null energy function and the Euler-Lagrange equation for $z$ applied to (\ref{plL2}) yield
\begin{eqnarray} \label{fe1a}
3\frac{\dot a^2}{a^2}=\frac{\rho_m^0}{a^3}+\frac{\rho_r^0}{a^4}+\frac{\beta_0^2\dot z^2}{2a^2}+\Lambda,\qquad
\dot z+\frac{M_0}{\beta_0^2}\frac{1}{a}=0,
\end{eqnarray}
respectively, which is a much simpler system then (\ref{sysa})-(\ref{sysc}) for $A=1/2$ and $B=0$.

From (\ref{fe1a}) it follows a differential equation just for $a$
\begin{equation}
3\frac{\dot a^2}{a^2}=\frac{\rho_m^0}{a^3}+\Bigg(\rho_r^0+\frac{M_0^2}{2\beta_0^2}\Bigg)\frac{1}{a^4}+\Lambda. \label{feN}
\end{equation}
This differential equation presents the same form of the usual Friedmann equation for matter, radiation and cosmological constant. Then, by neglecting the second term on the right-hand side of (\ref{feN}),  the known dominated matter -- cosmological constant solution can be obtained
\begin{equation}
a(t)={\Bigg(\frac{\rho^0_m}{\Lambda}\Bigg)}^{1\over3}\Bigg\{\sinh\Bigg[\frac{\sqrt{3\Lambda}}{2}(t+C_0)
\Bigg]\Bigg\}^{2\over3}, \label{sol1}
\end{equation}
where $C_0$ is a constant.
This limiting case corresponds to $\rho^0_r=0$ and $M_0=0$, then from (\ref{fe1a})$_2$ we have $z=\textrm{constant}$, which for the original variables implies the following solution for $\phi(t)={\phi_0}/{a(t)}$
\begin{equation}
\phi(t)=\phi_0{\Bigg(\frac{\Lambda}{\rho^0_m}\Bigg)}^{1\over3}\Bigg\{\sinh\Bigg[\frac{\sqrt{3\Lambda}}{2}(t+C_0)
\Bigg]\Bigg\}^{-{2\over3}}, \label{sola}
\end{equation}
where $\phi_0$ is a constant. Further, note that
$\lim_{t\rightarrow\infty}\phi(t)=0,$ so that $\lim_{t\rightarrow\infty}L(\phi)=1/2$,
which is according to the physical requirement that $L\longrightarrow 1/2$ at the present time.

On the other hand, when the radiation term dominates the matter term, the solution of (\ref{feN}) is
\begin{equation}
a(t)={\Bigg(\frac{2\beta_0^2\rho_r^0+M_0^2}{2\beta_0^2\Lambda}\Bigg)}^{1\over4}
\sqrt{\sinh\Bigg[2\sqrt{\frac{\Lambda}{3}}(t+C_1)\Bigg]}, \label{sol2}
\end{equation}
where $C_1$ is a constant. This case presents an additional initial energy density of radiation $\rho_0={M_0^2}/{2\beta_0^2}$, which is related to the scalar field non-minimally coupled. From the point of view of (\ref{feN}), this could be also interpreted as equivalent to the kinetic term of a scalar field minimally coupled to the gravity that evolves with $a^{-4}$, or simply as an additional term of radiation incorporated in $\rho_r$ -- $\rho_r=\rho_r^0a^{-4}\longrightarrow \rho_r=\left(\rho_r^0+M_0^2/2\beta_0^2\right)a^{-4}$ -- once they are indistinguishable from each other. Such an additional radiation term only appears in the early times if we admit $M_0\neq0$. This generates a radiation term greater than that assumed in the action of the model. Since we do not know if this extra radiation is detectable or not, it is a kind of dark radiation embedded in our $\Lambda$CDM model, when $M_0\neq0$. In a theoretical background, a type of dark radiation also appears in the brane-world cosmology \cite{13.00}. However, the recent cosmological data showed hints for an extra radiation field that could be related to neutrinos \cite{13.01}. Then it is possible to exist such an extra radiation. But note that from our model we cannot see the physical origin of this radiation, as it is just the result of the presence of a scalar field.

By using (\ref{fe1a})$_2$ and the solution (\ref{sol2}), we get
\begin{equation}
z=C_2-\frac{M_0}{\beta_0^2}{\Bigg(\frac{2\beta_0^2\Lambda}{2\beta_0^2\rho_r^0+M_0^2}\Bigg)}^{1\over4}
\!\int\!{\frac{dt}{\sqrt{\sinh\Big[2\sqrt{\frac{\Lambda}{3}}(t+C_1)\Big]}}}, \label{solb}
\end{equation}
with $C_2$ being a constant. Here we identify an elliptic integral of first kind, which furnishes the result
\begin{equation}
z=C_2-\frac{M_0}{2\beta_0^2}\sqrt{\frac{3}{\Lambda}}{\Bigg(\frac{2\beta_0^2\Lambda}{2\beta_0^2\rho_r^0+M_0^2}\Bigg)}^{1\over4}
F\bigg(\eta, \frac{1}{\sqrt{2}}\bigg), \label{solb1}
\end{equation}
where
\begin{equation}
\eta=\arccos{\Bigg\{\frac{1-\sinh\big[2\sqrt{\frac{\Lambda}{3}}(t+C_1)\big]}
{1+\sinh\big[2\sqrt{\frac{\Lambda}{3}}(t+C_1)\big]}\Bigg\}}. \label{solb2}
\end{equation}
Finally, we can obtain $\phi(t)$ directly from $z=a\phi/\beta_0$
\begin{equation}
\phi(t)=\phi_1\frac{F\Big(\eta, \frac{1}{\sqrt{2}}\Big)+C_3}{\sqrt{\sinh\Big[2\sqrt{\frac{\Lambda}{3}}(t+C_1)\Big]}},\label{scalarf}
\end{equation}
with
\begin{eqnarray}
\phi_1=-\frac{\sqrt{3}M_0}{\sqrt{2\Big(2\beta_0^2\rho_r^0+M_0^2\Big)}}, \qquad
C_3=-\frac{2\beta_0^2}{M_0}\sqrt{\frac{\Lambda}{3}}{\Bigg(\frac{2\beta_0^2\rho_r^0+M_0^2}{2\beta_0^2\Lambda}
\Bigg)}^{1\over4}C_2.
\end{eqnarray}

We observe that if the constant of motion $M_0$ vanishes in equation (\ref{feN}), its result is identical to that of the $\Lambda$CDM model (for the present era) and the solution $\phi=\phi_0/a$ is valid for the whole period that comprehends the radiation, matter and cosmological constant eras. It is interesting to note that starting from a Lagrangian very different from that of the cosmological constant model, the final dynamics is the same. The Lagrangian (\ref{plL1}) describes a Universe with a variable gravitational "constant", which presents a very different physics from that of the $\Lambda$CDM model. Further, it is important to saliently state here that the gravitational theory taken as the physical one is that given by the action (\ref{ga}), and the differential equation (\ref{feN}) -- with a general form as that from the cosmological constant model -- is the result of a transformation of coordinates in the configuration space which represents the dynamics that comes from (\ref{ga}) and not from the General Relativity plus cosmological constant. So we must naturally expect to detect some deviations of the present model from the $\Lambda$CDM model, as we will see below.

If we set  $\rho_r^0=0$ and $\rho_m^0=0$ in (\ref{feN}), one has a Universe only composed of scalar field non-minimally coupled, and the behavior of the scale factor will be
\begin{equation}
a(t)={\Bigg(\frac{M_0^2}{2\beta_0^2\Lambda}\Bigg)}^{1\over4}
\sqrt{\sinh\Bigg[2\sqrt{\frac{\Lambda}{3}}(t+C_1')\Bigg]}, \label{sol3}
\end{equation}
with $C_1'$ being a constant. Such a solution shows an expansion inevitably decelerated-accelerated, tending to a De Sitter Universe when $t\longrightarrow\infty$. From this, we infer that the presence of the radiation and matter fields will just retard the moment of the accelerated regime. On the other hand, a Universe filled only by a cosmological constant presents a genuine De Sitter expansion, the presence of matter and radiation being responsible for decelerated regimes.

Now it is interesting to extract some information on the dynamical system from the constant of motion generated by the Noether symmetry condition. From equation $(\ref{sysc})$ (for $B=0$), we obtain an expression with $\dot\phi$ directly related to the conserved quantity in the form
\begin{eqnarray}
\dot\phi=-\frac{M_0}{\beta_0a^2}-\frac{\dot a}{a}\phi. \label{icon}
\end{eqnarray}
Since one has a relation between the constant of motion and $\dot\phi$, we can link the constant of motion to the kinetic term of the scalar field. It is useful here to analyze the general dynamics by means of the energy density related to the scalar field. Applying (\ref{ed}) to the Noether forms, we rewrite the energy density of the scalar field in terms of the constant of motion by using (\ref{icon}) as
\begin{eqnarray}
&\rho_\phi=\frac{1}{2}\left[\frac{M_0^2}{\beta_0^2a^4}-\left(\frac{\dot a}{a}\right)^2\phi^2\right]+\Lambda.\label{icon1}
\label{ed1}
\end{eqnarray}
 Note that the above solutions show that $\phi$ stabilizes at a very small value in comparison to 1/2 independently on the initial condition of $\phi$, always leading to $L\longrightarrow1/2$ (General Relativity coupling). Taking account the mentioned behavior of $\phi$ and a large $a$, by looking at the equation (\ref{icon1}) one observes that the energy density of the scalar field valuated today tends to $\rho_\phi=\Lambda$. Analogously, from (\ref{pre}) and(\ref{icon}), we conclude that the pressure of the scalar field valuated today tends to $p_\phi=-\Lambda$. This behavior at the present time is clearly the cosmological constant-type one, which was observed in the solutions above obtained. Such a limiting solution exists because equation (\ref{icon}) states a relation that dictates that if $\phi$ dilutes then $\dot\phi$ also does. The deviation from the $\Lambda$CDM model occurs when going back to the past and can be seen by looking at the first term of (\ref{icon1}).

 Equation (\ref{icon}) indicates that the initial condition of $\dot\phi$ could be directly related to the conserved quantity in the form
\begin{eqnarray}
\dot\phi(0)=-\frac{M_0}{\beta_0}-H(0)\phi(0), \label{icon2}
\end{eqnarray}
where $H(0)$ and $\phi(0)$ are the Hubble parameter ($H=\dot a/a$) and the scalar field at $t=0$, immediately after inflation (say). The scale factor was normalized as $a(0)=1$ for cleanness of the analysis. From the coupling $L=\frac{1}{2}(1-\frac{\phi^2}{6})$ one determines the possible range of values for $\phi(0)$ by the requirement that $L>0$, which guarantees attractive gravity, giving $-\sqrt{6}<\phi(0)<+\sqrt{6}$. These constraints on the initial condition of $\phi$ together with the requirement of the weak energy condition at $t=0$, $\rho_\phi(0)\geq0$, could constrain the deviation from the $\Lambda$CDM model in the past. As an example, we may take the simpler case $M_0=0$, when $\phi$ dilutes with the inverse of $a$. In so doing, the weak energy condition gives the constraint
\begin{eqnarray}
-\sqrt{6}<\phi(0)\leq\frac{\sqrt{2\Lambda}}{H(0)}\quad\Longrightarrow\quad-\sqrt{2\Lambda}\leq\dot\phi(0)<H(0)\sqrt{6},
\end{eqnarray}
which limits the possible deviations of the present model for a vanishing constant of motion from the cosmological constant model. For this case, since $\phi$ dilutes with time, we must just fix its adequate initial condition and the system will consistently evolve in all the eras. This is, the weak energy condition is respected and the coupling never reaches the value $L(\phi)=0$.

The evolution of the scalar field in the early times for the case $M_0\neq 0$, solution (\ref{scalarf}), is not simple and it is desirable assuring that the scalar field always dilutes with time, as for the case $M_0=0$, avoiding any possible inconsistency after fixing the initial conditions. In order to know about its behavior, we rewrite equation (\ref{scalarf}) in terms of the parameter $\eta$ through (\ref{solb2}). Further, we can set $C_2=0$ (which implies $C_3=0$) because $\phi$ dilutes with the inverse of $a$ only for large $t$ -- the term with the elliptic function dominates in the early times. In so doing, one obtains
\begin{eqnarray}
\frac{\phi(\eta)}{\phi_1}=\sqrt{\frac{1+\cos{\eta}}{1-\cos{\eta}}}F\left(\eta, \frac{1}{\sqrt{2}}\right).
\end{eqnarray}
In this expression, an increase in $\eta$ corresponds to an increase in time. The corresponding behavior is such that $\phi/\phi_1$ slowly increases until a maximum at $\eta\approx 0.44 \pi$ and after it continuously decreases. Thus $\phi$ presents the adequate behavior at any point $\eta_0> 0.44 \pi$. By considering this constraint and using (\ref{solb2}) to establish that $\eta=\eta_0$ at $t=0$, we have that the constant $C_1$ must obey the relation
 \begin{eqnarray}
C_1=\frac{1}{2}\sqrt{\frac{3}{\Lambda}}\textrm{arcsinh}\left(\frac{1-\cos{\eta_0}}{1+\cos{\eta_0}}\right).
\end{eqnarray}
The constant $C_1$ given in this form assures that $\phi/\phi_1$ is a decreasing function of time. Taking the above considerations, we can now guarantee the weak energy condition and avoid the vanishing of the coupling (i.e. $\phi^2=6$) by fixing the adequate initial condition of $\phi$, as it is done for the case $M_0=0$.

Since the dilution of $\phi$ with time is guaranteed, we can also constrain the variation of the gravitational constant from the constraint on the initial condition of $\phi$. This consists of a necessary refinement of the previous constraints on the initial conditions. The effective gravitational constant begins with a higher value and decreases until it stabilizes in the late times. Therefore we must fix the initial condition of $\phi$ in accordance with the observations and then the time evolution of $L(\phi)$ will not violate the observational data. Firstly we consider that in the primordial Universe the gravitational constant was $G=(1+\delta)G_0$, with $G_0$ being its value today. The quantity $\delta$ determines how much bigger the gravitational constant was in the past. Taking a relatively small $\delta$, as the observations suggest, the gravitational coupling $L(\phi)$ can be valuated at the initial instant as follows
\begin{eqnarray}
L[\phi(0)]=\frac{1}{2}\left[1-\frac{\phi(0)^2}{6}\right]=\frac{1}{2(1+\delta)}\approx\frac{1}{2}(1-\delta+\delta^2),
\end{eqnarray}
from which we immediately determine the initial condition of $\phi$, namely,
\begin{eqnarray}
\phi(0)^2={6(\delta-\delta^2)}.
\end{eqnarray}
This initial condition will assure that the maximum value of the gravitational constant in the past is in accordance with the observations. The work \cite{12} establishes $-0.10<\delta<0.13$ for the period comprehending BBN and today, and reference \cite{13} establishes $-0.083<\delta<0.095$ for the period between recombination and today.

Let us now compare our results with other works. The paper \cite{10} analyzes the non-minimally coupled scalar field for a F-R-W metric in the absence of a radiation field. The model admits a family of forms for the potential and coupling satisfying the Noether symmetry condition. On the other hand, in our model we take account the radiation field, in a flat F-R-W metric ($k=0$), which strongly constrains the potential (for more details on the solutions for the case $k=0$ in the absence of radiation see references \cite{9.0, 10.1}). This constraining comes from equation (\ref{sys4}), which is affected by the radiation term. For a Noether symmetry satisfied by $\alpha$ and $\beta$ given as separable functions of $a$ and $\phi$, we found only one set of potential and coupling, i.e. a quadratic coupling and a constant potential. In the absence of a radiation field, the possibilities to the Noether potential and coupling are increased, which will coincide with those discussed in reference \cite{10}, for the case $k=0$. In the work \cite{11}, multiple scalar fields minimally coupled with gravity in the absence of a radiation field are studied, comprehending quintessence, phantom, and quintom models. The corresponding Noether potentials can take several forms, generating families of solutions according to each model. We can expect that such potentials would be strongly constrained in the presence of a radiation field, similarly to what happens for the case with the non-minimally coupled scalar field.

\section{Conclusions}

To sum up, we have investigated a non-minimally coupled scalar field model by using the Noether symmetry approach. A constant potential and a conformal coupling emerged from the Noether symmetry condition, such that the conformal coupling becomes related to a symmetry of the model. From these results, by means of a change of coordinates in the configuration space -- generated by the found symmetry --  we could reduce the original dynamical system. The final field equation was of the same form of the Friedmann equation for the cosmological constant model. Hence, the resulting dynamics was similar to that of the $\Lambda$CDM model. This mathematical artifice could mask the physics behind the non-minimally coupled scalar field and allow the $\Lambda$CDM model to emerge. Further, with this method we have shown that such a model with a non-minimal coupling, i.e. with a physics where the gravitational "constant"  is variable, can reproduce the observed dynamics from the radiation era until the accelerated era quite exactly as the $\Lambda$CDM model does. The constant of motion furnished information on the range of the possible initial conditions of the scalar field, which constrains the deviations of the theory from the cosmological constant model in the early Universe. As a refinement of the adequate initial conditions of the scalar field, we constrained the time evolution of the scalar field and derived an expression that constrains the initial condition of the scalar field such that the variation of the gravitational constant is in accordance with the observations.

\end{document}